\documentclass[final,3p,times,twocolumn,fleqn]{elsarticle}
\usepackage{amssymb}
\usepackage{graphicx}
\usepackage{amsmath}
\usepackage{subfigure}
\usepackage{epsfig}

\biboptions{compress}

\begin{document}
%\LARGE
\begin{frontmatter}
%\LARGE
\title{\bf Phase sensitive quantum interference on forbidden transition in ladder scheme}

\author{Gennady A. Koganov\corref{cor1}}
\ead{quant@bgu.ac.il}

\author{Reuben Shuker}
\ead{shuker@bgu.ac.il}

\cortext[cor1]{Corresponding author: Gennady Koganov}

\address{Physics Department, Ben-Gurion University of the Negev, \\ P.O.B. 653, Beer Sheva, 84105, Israel}

\begin{abstract}
A three level ladder system is analyzed and the coherence of initially electric-dipole forbidden transition is calculated. Due to the presence of two laser fields the initially dipole forbidden transition becomes dynamically permitted due to ac Stark effect. It is shown that such transitions exhibit quantum-interference-related phenomena, such as electromagnetically induced transparency, gain without inversion and
enhanced refractive index. Gain and dispersion characteristics of such transitions strongly depend upon the relative phase between the driving and the probe fields. Unlike allowed transitions, gain/absorption behavior of ac-Stark allowed transitions exhibit antisymmetric feature on the Rabi sidebands. It is found that absorption/gain spectra possess extremely narrow sub-natural resonances on these ac Stark allowed forbidden transitions. An interesting finding is simultaneous existence of gain and negative dispersion at Autler-Townes transition which may lead to both reduction of the group velocity and amplification of light.
\end{abstract}


\end{frontmatter}

%\maketitle
\section{Introduction}

\begin{figure}[htbp]
\centering
\includegraphics[scale=0.6]{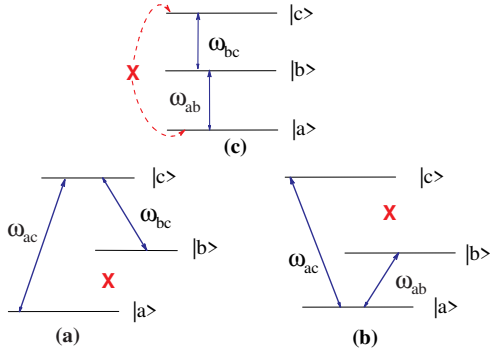}
\caption{Schematic $\Lambda$ (a), $V$ (b) and ladder (c) three level
configurations. Transitions
$\left|a\right\rangle\rightarrow\left|b\right\rangle$,
$\left|b\right\rangle\rightarrow\left|c\right\rangle$ and
$\left|a\right\rangle\rightarrow\left|c\right\rangle$ are initially dipole-forbidden (marked with red cross).} \label{Fig1}
\end{figure}

Phase sensitive detection is of major importance in physical quantity measurements. A number of works on phase sensitive phenomena, both theoretical and experimental have been published recently (see, e.g. Refs. \cite{Ruby,SelectionRules,Dynes-2006,Wilson-Gordon} and references therein). In this paper we investigate quantum control and interference in a three level ladder system. This approach can be extended to other schemes and generalized to more levels.
 High dependence and sensitivity on the relative phase of the pump and probe em modes of the relevant lasers is found and presented here. This sensitivity is particularly accentuated in the third transition, whose frequency is essentially the sum of the frequencies of the sequential transitions in the ladder, even when this transition is one photon emission forbidden in the bare state \cite{Forbidden}, i.e., without the presence of the pump and probe fields. Similarly to semiconductor structures \cite{SelectionRules,Frog-Nature,Dynes-2005,QuantDots}, where all three transitions in the ladder three level scheme are allowed due to dc Stark mixing, in atomic ladder system the initially forbidden transition becomes dynamically allowed due to ac Stark mixing. Recently experimental evidence of such an effect was reported \cite{Havey}. A most interesting finding is that the effect of such phase control renders amplification, absorption and dispersion at will on an initially unallowed transition. Moreover, subnatural linewidth resonances are found, which can be detuned from both Autler-Townes \cite{Autler} and bare resonances. Phase control provides a key for detection, amplification, enhanced dispersion as well as anomalous negative index of refraction, dark states and other quantum characteristics. Our findings are not limited to the ladder scheme but rather also applicable, with the proper modifications, to the $\Lambda$ and V schemes which are shown in Fig.\ref{Fig1}.
 
In order to provide physical intuition and understanding, electric dipole forbidden transition, becomes allowed by ac Stark mixing with other allowed transitions as a result of the strong electric field of the impinging electromagnetic radiation, as detailed below. Gain without inversion at the third transition in the ladder scheme is of particular interest in the X-Ray regime, where coherent sources are scarce. 

Quantum-interference-related phenomena, such as electromagnetically
induced transparency (EIT), lasing without inversion (ScullyBook) etc., are
based on the interference between two independent quantum channels
\cite{ScullyBook,EIT-review,LWI-review,Quantum-control,Velichan,Maser,Rb-maser}.
Traditional treatment of such phenomena typically involves a three
level scheme and two coherent fields, a strong driving field and a
weak probe one, applied to the two dipole-allowed transitions, followed
by measurement of the absorption and the dispersion of the probe
transition. 

We show in the following that dipole-forbidden transitions can also exhibit amplification, enhanced dispersion and other coherent phenomena. The electric field component of the driving laser field breaks the space spherical symmetry and renders the parity not well defined, as in \textit{dc} Stark effect. In other words, the presence of a strong driving field exerts \textit{ac} Stark effect and thus breaks the spherical symmetry of the system and creates an infinite ladder of dressed states \cite{CohenTanudji,Doron2001}, so that transitions
between the dressed states are not necessarily constrained by the
selection rules for a free atomic system. An interesting signature of the forbidden transitions is the antisymmetric character of the gain and dispersion on the Rabi sidebands. We found that gain and dispersion properties of such transitions are phase sensitive as they strongly depend upon the relative phase between the driving and probe fields. Our results open a perspective for new type of phase sensitive spectroscopy in a wide spectral range, from microwave to X-ray frequencies.

\section{Hamiltonian, master equation and dressed states}
There are three possible closed-loop three-level configurations,
$\Lambda$, V and the ladder schemes. These are shown in Fig.
\ref{Fig1}. All of the three schemes possess a common property that due to selection rules, one of the three transitions is dipole-forbidden by parity which is well defined in the absence of the laser fields. However, the parity becomes ill defined due to the presence of the ac electric fields of the two impinging lasers. These fields break the space symmetry and mix states of different parity. Hence the originally forbidden transition becomes undefined. In this sense it becomes \textit{dynamically} allowed. We call such dynamically allowed transitions ac-Stark allowed (ACSA) transitions. In a sense this is a
different variant than the usual EIT or LWI schemes as here the
transition is forbidden to begin with and there is no issue of
population inversion. Such a possibility would be particularly
important in the X-Ray regime for gain in a ladder scheme
\cite{Doron2003}. Other examples are forbidden transitions
in Alkali atoms.

\begin{figure}[htbp]
\centering
\includegraphics[scale=0.2,viewport=0 0 500 500,clip]{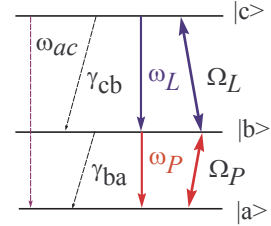}
\caption{Schematic  ladder  three level
configuration. Transitions
$\left|a\right\rangle\rightarrow\left|b\right\rangle$ and 
$\left|b\right\rangle\rightarrow\left|c\right\rangle$ are dipole allowed, while the transition 
$\left|a\right\rangle\rightarrow\left|c\right\rangle$ is initially dipole-forbidden. $\omega_{P}$/$\omega_{L}$ and $\Omega_{P}$/$\Omega_{L}$ are optical and Rabi frequencies of the probe/drive fields, respectively, $\omega_{ac}$ is the energy difference between the levels $\left|c\right\rangle$ and $\left|a\right\rangle$, not the transition frequency in the bare scheme. $\gamma_{ba}$ and $\gamma_{cb}$ are the rates of spontaneous emission on the two relevant allowed transitions} \label{Ladder}
\end{figure}

Without loss of generality among the
three above mentioned configurations, we consider the Ladder-configuration (see
Fig. \ref{Ladder}). A strong driving field with frequency $\omega_{L}$ is applied to atomic
transition $\left|b\right\rangle\rightarrow\left|c\right\rangle$, and a probe field with frequency $\omega_{P}$ is applied to atomic transition $\left|a\right\rangle\rightarrow\left|b\right\rangle$. The time dependent interaction Hamiltonian in the interaction picture is given by

\begin{equation}\label{Hamiltonian}
H_{int}=\hbar\Omega_{P}e^{i(\Delta_{P}t+\varphi_{P})}\sigma_{ab}+\Omega_{L}e^{i(\Delta_{L}t+\varphi_{L})}\sigma_{bc} + H.c.
\end{equation}

\noindent where $\sigma_{ab}=\left|b\rangle\langle a\right|$, $\sigma_{bc}=\left|c\rangle\langle b\right|$, and $\sigma_{ac}=\left|c\rangle\langle a\right|$ are the atomic rising operators,  $\Omega_{L}$ and $\Omega_{P}$ are Rabi frequencies of the driving and the probe fields, respectively, $\Delta_{L}$ and $\Delta_{P}$ are detunings of the fields from corresponding atomic transitions. We emphasize the introduction of the phases $\varphi_{L}$ and $\varphi_{P}$  of the driving and the probe fields introduced explicitly in the Hamiltonian. The master equation is given by

\begin{equation}\label{master}
\frac{\partial \rho}{\partial t}=-\frac{i}{\hbar}[H_{int},\rho]+ 
\end{equation} 
\begin{equation*}
\frac{1}{2}[\gamma_{ba}(2\sigma_{ba}\rho\sigma_{ab}-\sigma_{ab}\sigma_{ba}\rho)+\gamma_{cb}(2\sigma_{cb}\rho\sigma_{bc}-\sigma_{bc}\sigma_{cb}\rho)]
\end{equation*}

\begin{figure}[htbp]
\begin{center}
\includegraphics[scale=0.6]{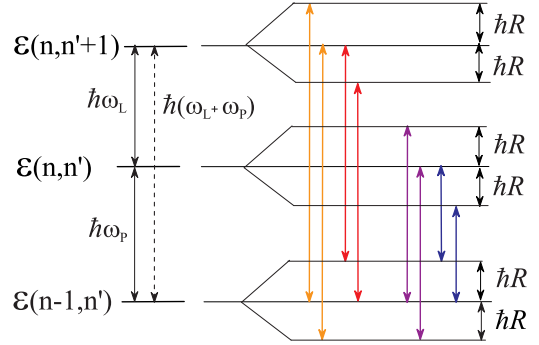}
\end{center}
\caption{Dressed states picture at bare two-photon resonance
$\Delta_{P}+\Delta_{L}=0$. On the left: manifold of bare states labeled
by the atomic level index \textit{a}, \textit{b}, or \textit{c},
the probe field photon number \textit{n}, and the driving field
photon number \textit{n'}.  On the right: dressed states of coupled
atom+field system. Strong phase sensitive gain
without inversion is obtained on the ACSA transition
$\left|a\right\rangle\rightarrow\left|c\right\rangle$ at
frequencies $\omega_{P}+\omega_{L}+R$, left couple of thick arrows (orange on line) and $\omega_{P}+\omega_{L}-R$, right couple of thick arrows (red on line). Two couples of thin arrows(blue and violet on line), show the probe transition frequencies $\omega_{P}-R$ and
$\omega_{P}+R$, where gain is also possible.
$R=\sqrt{\Omega_{P}^{2}+\Omega_{L}^{2}}$ is the generalized Rabi
frequency.} \label{DressedStates}
\end{figure}

To analyze the steady state we eliminate the time dependence using the rotating wave approximation, then the system Hamiltonian is given by

\begin{equation}\label{Steady-Hamiltonian}
H=\left(
\begin{array}{lll}
 0 & \Omega_{L}e^{i\varphi_{L}} & 0 \\
\Omega_{L}e^{-i\varphi_{L}} & \Delta_{L} & \Omega_{P}e^{i\varphi_{P}} \\
0 & \Omega_{P}e^{-i\varphi_{P}} & \Delta_{L}+\Delta_{P}
\end{array}
\right)
\end{equation}

\noindent where $\Omega_{P}$ and $\Omega_{L}$ are Rabi frequencies of the probe and the driving fields, respectively, and $\Delta_{P}$ and $\Delta_{L}$ are corresponding detunings.  Abbreviated manifold of dressed states created by the strong driving field $\Omega_{L}$ is
shown in Fig. \ref{DressedStates}. The corresponding semiclassical dressed states in the case of two-photon resonance, i.e. $\Delta_{P}+\Delta_{L}=0$ are given by

\begin{equation}
\left|\pm\right\rangle=\frac{1}{\sqrt{2}}(\frac{\Omega_{P}}{\sqrt{\Omega_{L}^{2}+\Omega_{P}^{2}}}\left|a\right\rangle\pm e^{-i\varphi_{P}}\left|b\right\rangle +\frac{e^{-i(\varphi_{P}+\varphi_{L})}\Omega_{L}}{\sqrt{\Omega_{L}^{2}+\Omega_{P}^{2}}} \left|c\right\rangle)
\end{equation}
 
\noindent These dressed states are superposition of states of different parity and hence are not constrained by the selection rules for atomic states $\left|a\right\rangle, \left|b\right\rangle$, and $\left|c\right\rangle$. As will be seen in the following,
maximal gain is achieved when the probe and the drive lasers are in
''dressed'' two-photon resonance with transitions between the
dressed states marked with red and orange arrows, i.e. at
$\Delta_{P}+\Delta_{L}=\pm R$, where $R=\sqrt{\Omega_{P}^{2}+\Omega_{L}^{2}}$ is the generalized Rabi frequency.\\

\section{Steady state coherences}
We have solved analytically master equation (\ref{master}) 
for the atomic density matrix $\rho$ with the Hamiltonian (\ref{Steady-Hamiltonian}) in steady state and calculated the coherences $\rho_{ab}$ and $\rho_{ac}$ on the probe and the ACSA transitions, whose imaginary and real parts are related to absorption/gain and
dispersion, respectively. The main features can be qualitatively
understood from the approximate expressions for the coherences
$\rho_{ab}$ and $\rho_{ac}$ on probe and ACSA transitions at
small probe field $\Omega_{P}\ll\Omega_{L}$, although all graphical
results presented in the following have been obtained from the
exact analytic solution of the steady state master equation. 
Approximate formulas for the coherences are given by

\begin{equation}
\rho_{ab}= -\frac{2\Omega_{P}e^{i\varphi_{P}}[2(\Delta_{L}+\Delta_{P})+i\gamma_{cb}]}{(\gamma_{ba}-2i\Delta_{P})[\gamma_{cb}-2i(\Delta_{L}+\Delta_{P})]+4\Omega_{L}^{2}}\label{r01}
\end{equation}
\begin{equation}
\rho_{ac}=-\frac{4\Omega_{P}\Omega_{L}e^{i(\varphi_{P}+\varphi_{L})}}{(\gamma_{ba}-2i\Delta_{P})[\gamma_{cb}-2i(\Delta_{L}+\Delta_{P})]+4\Omega_{L}^{2}}\label{r02}
\end{equation}

\noindent Here $\gamma_{cb}$ and $\gamma_{ba}$ are spontaneous decay
rates on the drive and the probe transitions, respectively,
$\Delta_{P}=\omega_{ab}-\omega_{P}$ and
$\Delta_{L}=\omega_{bc}-\omega_{L}$ are one-photon detunings between
the laser frequencies $\omega_{P}$ and $\omega_{L}$ and the
corresponding atomic frequencies $\omega_{ab}$ and $\omega_{bc}$.
The two-photon detuning
$\Delta_{P}+\Delta_{L}=\omega_{ac}-(\omega_{P}+\omega_{L})$ contains
the frequency $\omega_{ac}$ of originally dipole-forbidden atomic
transition $\left|a\right\rangle\rightarrow\left|c\right\rangle$
indicating possible oscillations at that frequency. It is important to note that
due to the presence of the exponent factor
$exp[i(\varphi_{P}+\varphi_{L})]$ in the numerator of Eq.
(\ref{r02}), the coherence $\rho_{ac}$ is \textit{phase sensitive}. Varying the relative phase
$\Delta\varphi=\varphi_{P}+\varphi_{L}$ between the probe and
 the drive fields
allows for an interchange between absorptive and dispersive line shapes of
$\left|a\right\rangle\rightarrow\left|c\right\rangle$ transition, defined by imaginary and real parts of the coherence $\rho_{ac}$,
respectively. It is instructive to consider two particular cases
of "bare" and "dressed" two-photon resonances, when
$\Delta_{P}+\Delta_{L}=0$ and $\Delta_{L}+\Delta_{P}=\pm
\Omega_{L}$, respectively. At "bare" two-photon resonance $\Delta_{P}+\Delta_{L}=0$, the coherence $\rho_{ab}$ on the probe transition is small of the order of $\Omega_{P}/\Omega_{L}^{2}$. the coherence $\rho_{ac}$ on the ACSA transition is proportional to $e^{i\Delta\varphi}\Omega_{P}/\Omega_{L}$, so that both absorption and dispersion can take
small values between $-\Omega_{P}/\Omega_{L}$ and
$\Omega_{P}/\Omega_{L}$, depending on the relative phase $\Delta\varphi$. This accentuates the importance of the phase relation between the two laser fields and the phase sensitivity of the ACSA transition.

A different picture is obtained when the probe field is tuned in
resonance with transitions between dressed states, i.e. when the two
photon detuning equals the Rabi frequency of the driving field ("dressed" two-photon
resonance). The dressed resonance condition and the coherences are given by

\begin{equation}
\Delta_{L}+\Delta_{P}=\pm \Omega_{L} \label{dressed-res}
\end{equation}
\begin{equation}
\rho_{ab}= \pm\frac{2e^{i\varphi_{P}}(\gamma_{cb}\mp 2i\Omega_{L})\Omega_{P}}{2(\gamma_{ba}+\gamma_{cb})\Omega_{L}\pm i\gamma_{ba}\gamma_{cb}}\label{r01-res-dress}
\end{equation}
\begin{equation}
\rho_{ac}= \mp\frac{ie^{i(\varphi_{P}+\varphi_{L})}\Omega_{L}\Omega_{P}}{2(\gamma_{ba}+\gamma_{cb})\Omega_{L}\pm i\gamma_{ba}\gamma_{cb}}\label{r02-res-dress}
\end{equation}

\noindent In this case both $\rho_{ac}$ and $\rho_{ab}$ are not small since they
are proportional to the probe Rabi frequency $\Omega_{P}$ rather
than to the ratio $\Omega_{P}/\Omega_{L}$ between the probe and the
driving field Rabi frequencies, as in the above mentioned bare resonance case. This is confirmed by the exact analytical calculations as discussed below. Also gain and dispersion on the ACSA
transition are not symmetric with respect to the bare two-photon
resonance $\Delta_{P}+\Delta_{L}=0$, as evident from Eq. (\ref{r02-res-dress}). \\

\begin{figure}[htbp]
%\centering
%\includegraphics[scale=0.55]{Fig5.eps}
\subfigure[]{\label{DFi=0}\includegraphics[scale=0.3]{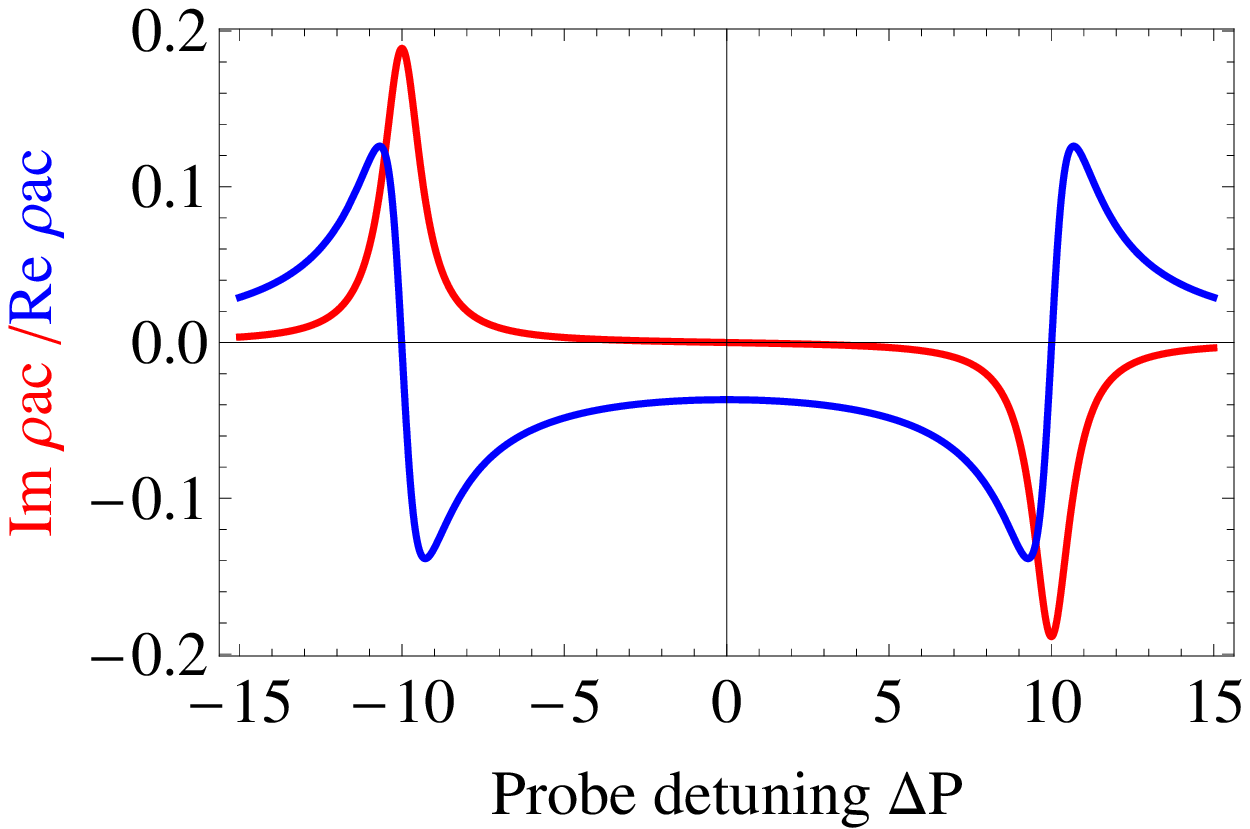}}
\subfigure[]{\label{DFi=05Pi}\includegraphics[scale=0.3]{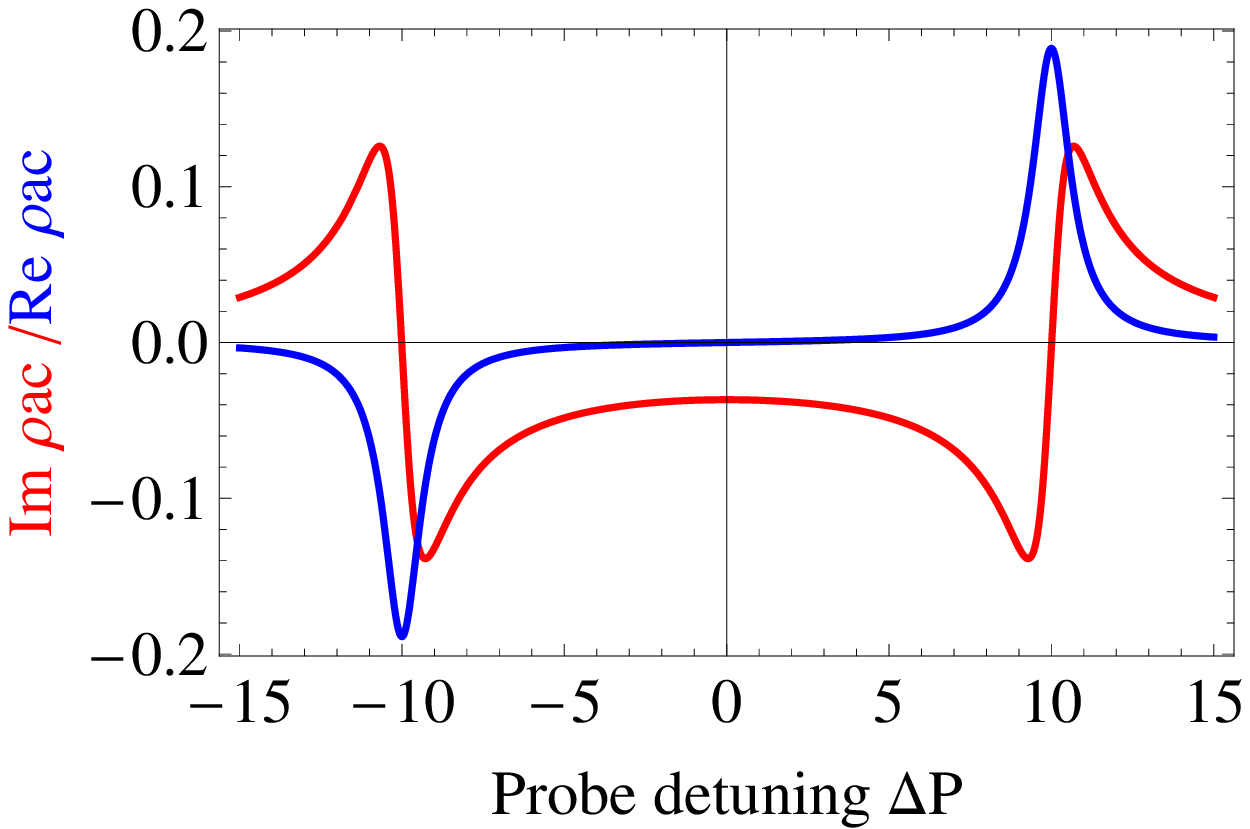}}\\
\subfigure[]{\label{DFi=Pi}\includegraphics[scale=0.3]{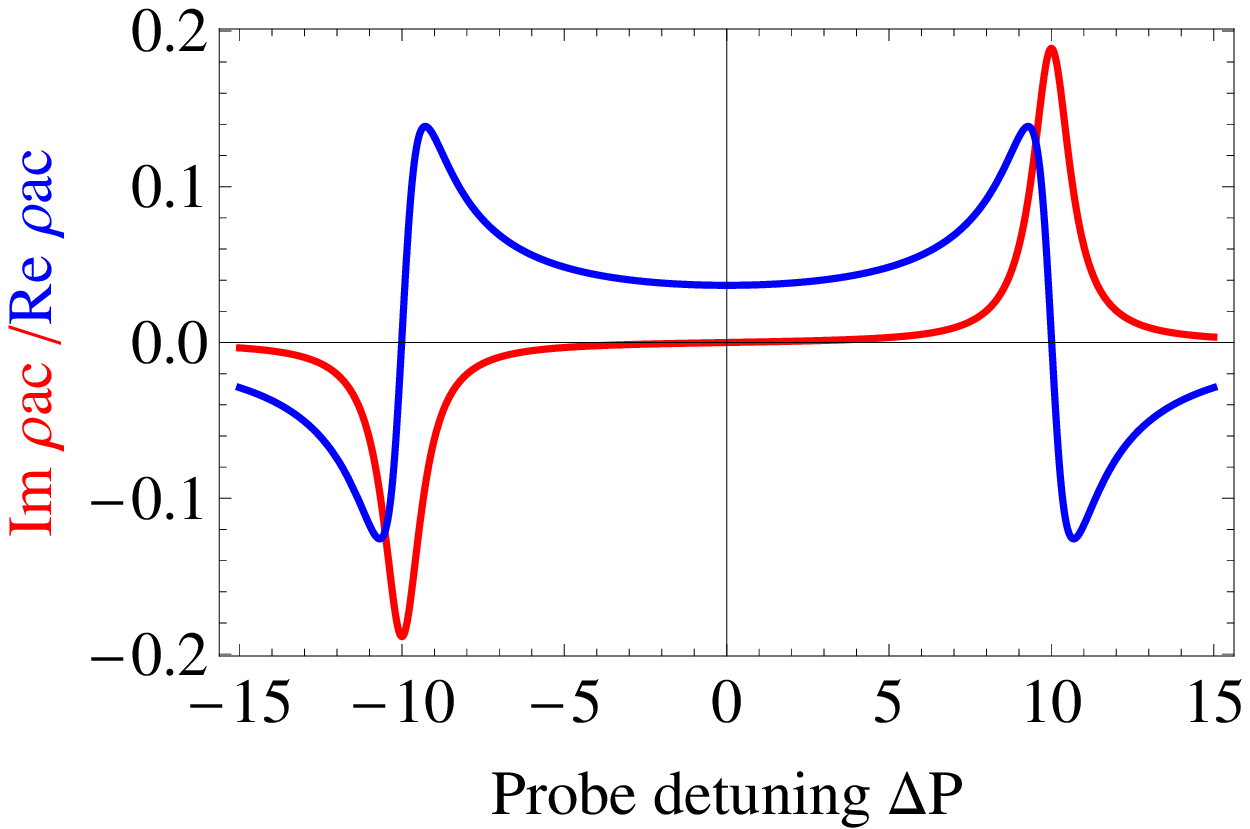}}
%\subfigure[]{\label{r02=-5}\includegraphics[scale=0.23]{Lam-Forb-Spectr-dP=-5.eps}}
\subfigure[]{\label{DFi=15Pi}\includegraphics[scale=0.3]{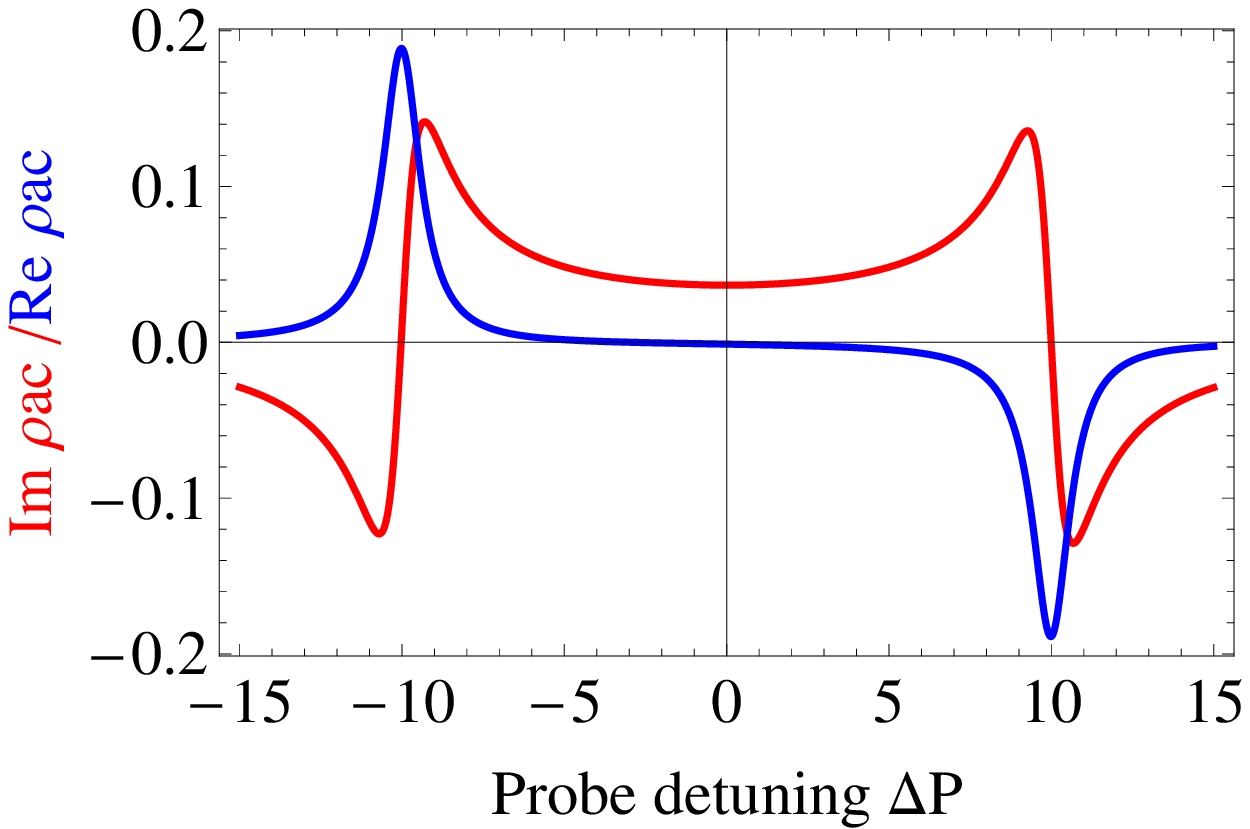}}
\caption{Gain (red line) and dispersion (blue line) at ACSA transition as a function of the probe detuning. Parameters: $\gamma_{ba}=\gamma_{cb}$,
$\Omega_{L}=10\gamma_{cb}$, $\Omega_{P}=0.37\gamma_{cb}$, $\Delta\varphi=0$ (a),  $\pi/2$ (b),  $\pi$ (c), $3\pi/2$ (d).}\label{animation1}
\end{figure}

\begin{figure}[htbp]
\centering
\subfigure[]{\label{Gain}\includegraphics[scale=0.25]{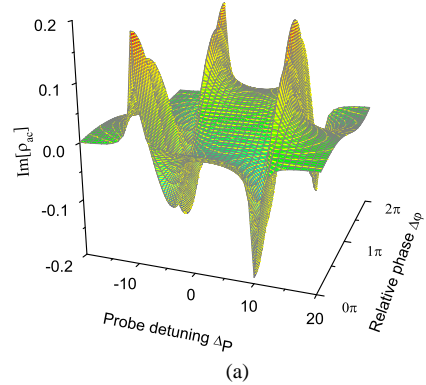}}\\
\subfigure[]{\label{Index}\includegraphics[scale=0.25]{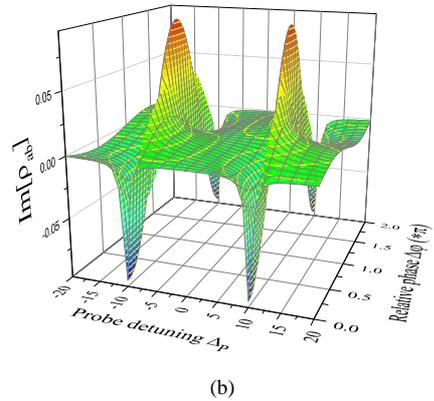}}
\caption{3D plot of gain/absorption at ACSA (a) and probe (b) transition as a function of the probe detuning and of the relative field phase $\Delta\varphi$. Parameters: $\gamma_{ba}=\gamma_{cb}$,
$\Omega_{L}=10\gamma_{cb}$, $\Omega_{P}=0.37\gamma_{cb}$.}\label{Ladder-Forbid-3D}
\end{figure}

To get a quantitative picture, the results of the exact analytical solution of the steady state master equation are shown in the following. Steady state coherence $\rho_{ac}$ on the ACSA transition as a function of the probe detuning is shown in Fig. \ref{animation1} at various values of the relative phase $\Delta\varphi$. 

The coherence $\rho_{ab}$ is drawn in 3D Fig. \ref{Ladder-Forbid-3D} as
a function of the probe detuning $\Delta_{P}$ and of the relative field phase $\Delta\varphi$.
When the probe and the driving fields are \textit{in-phase} ($\Delta\varphi=0$), there is a strong gain with negative dispersion slope at $\Delta_{P}+\Delta_{L}=-\Omega_{L}$, and
absorption with normal dispersion at $\Delta_{P}+\Delta_{L}=\Omega_{L}$.
Inverse picture is obtained when the probe and driving fields are
\textit{out of phase}, i.e. $\Delta\varphi=\pi$. If the field phases are
$\pm\pi/2$-shifted, the absorption profile takes a
dispersive shape while the dispersion behaves in absorptive-like
manner - an indication of the quantum interference. Note the
difference between the two transitions: the probe absorption $Im[\rho_{ab}]$ is symmetric with respect to bare two-photon resonance $\Delta_{P}+\Delta_{L}=0$ so that
there is either absorption or gain simultaneously on both side-bands
$\Delta_{P}+\Delta_{L}=\pm\Omega_{L}$, while the coherence $Im[\rho_{ac}]$ on the ACSA transition is
antisymmetric with respect to the origin, so that there is always gain on one side-band and
absorption on the other one.
The frequency at which amplification takes place can be
tuned by varying the driving field intensity because maximal gain
is obtained at $\Delta_{P}+\Delta_{L}=\pm\Omega_{L}$.

\begin{figure}[htbp]
\centering
\subfigure[]{\label{FlatGain}\includegraphics[scale=0.3]{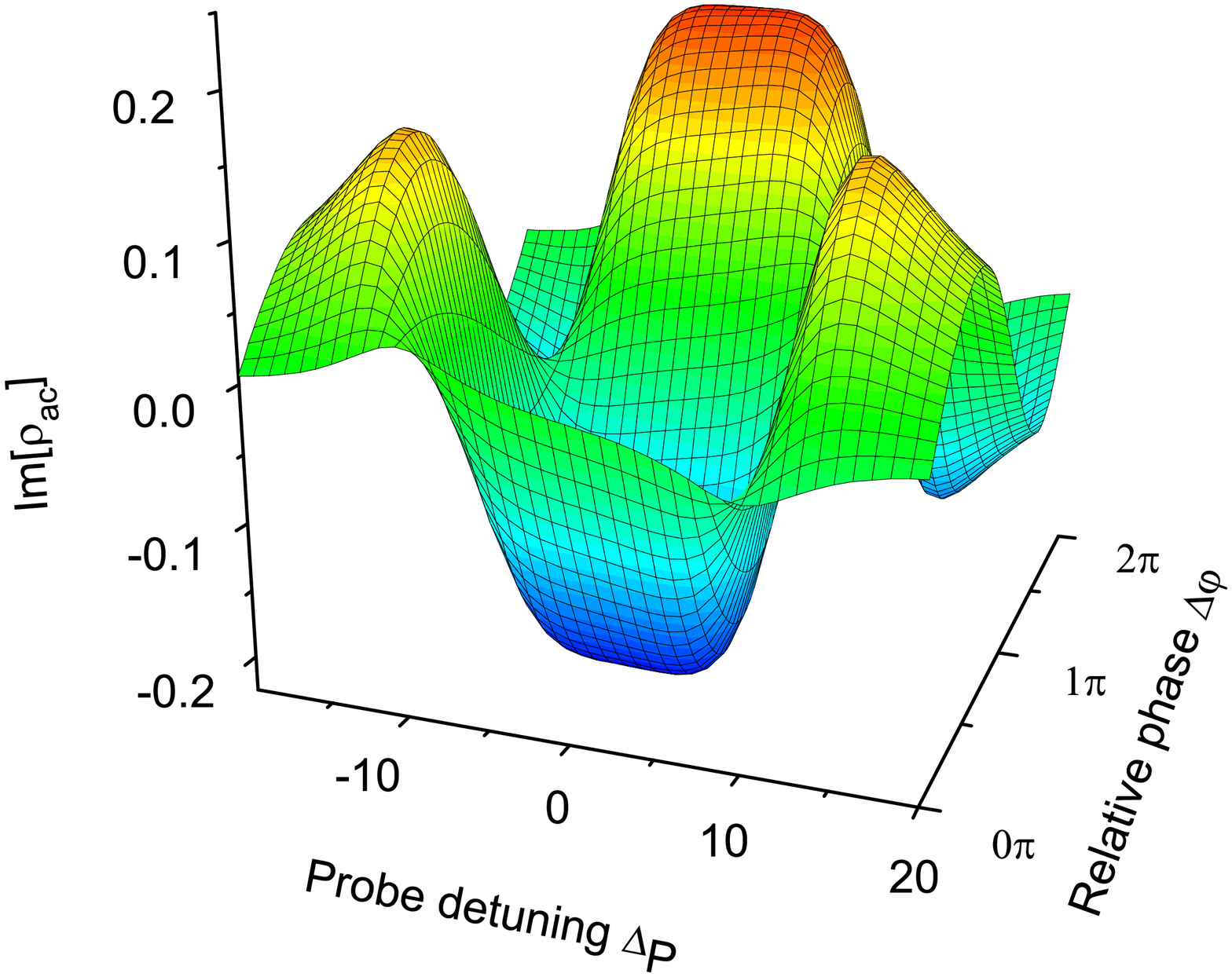}}
\subfigure[]{\label{FlatIndex}\includegraphics[scale=0.3]{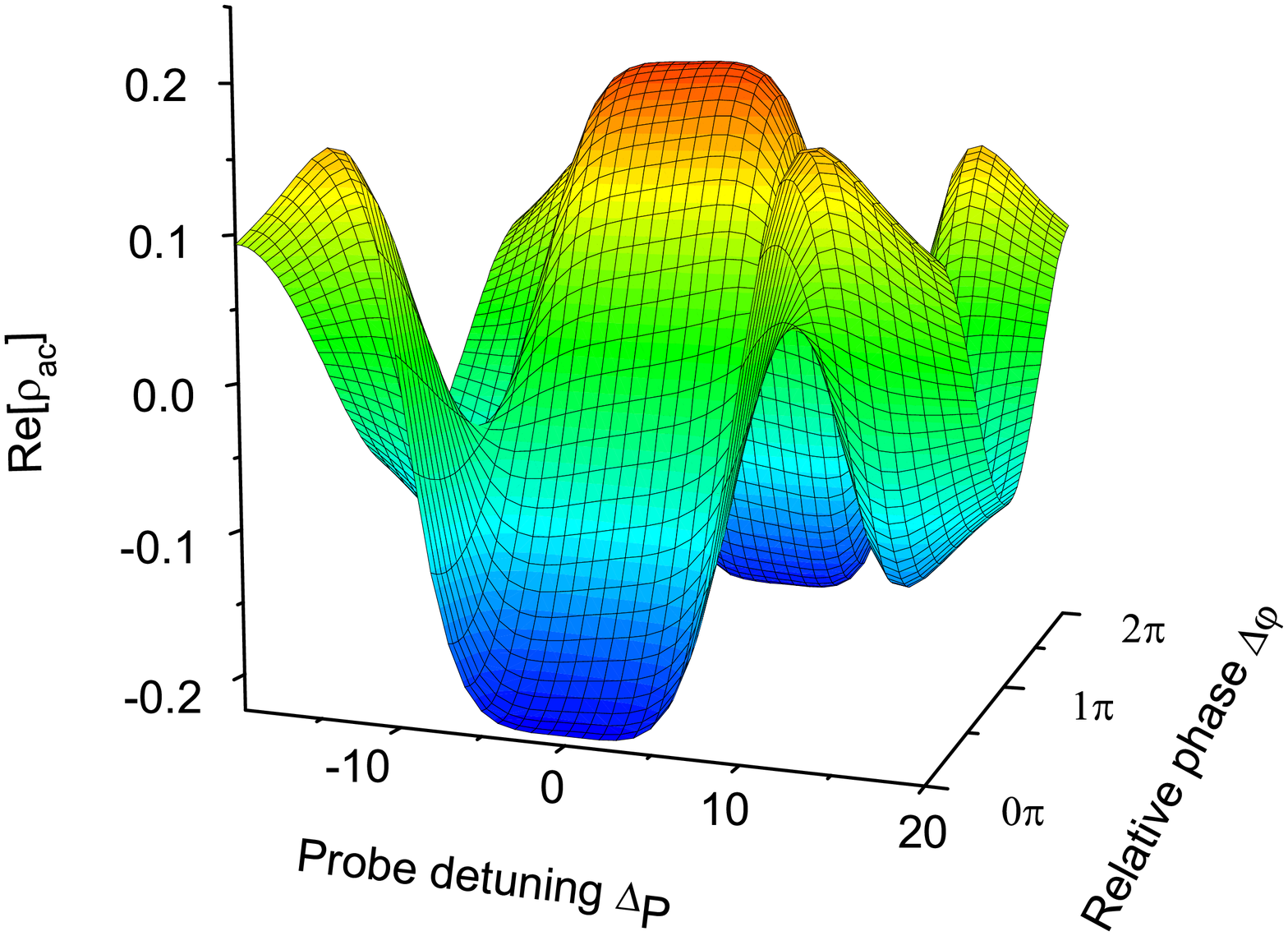}}
\caption{3D plot of gain/absorption (a) and dispersion (b) at ACSA transition at strong probe field as a function of the probe detuning and of the relative field phase $\Delta\varphi$. Parameters: $\gamma_{ba}=\gamma_{cb}$,
$\Omega_{L}=10\gamma_{cb}$, $\Omega_{P}=3.4\gamma_{cb}$.}
\label{Flattop-3D}
\end{figure}

\begin{figure}[htbp]
\centering
\subfigure[]{\label{DFi=0}\includegraphics[scale=0.3]{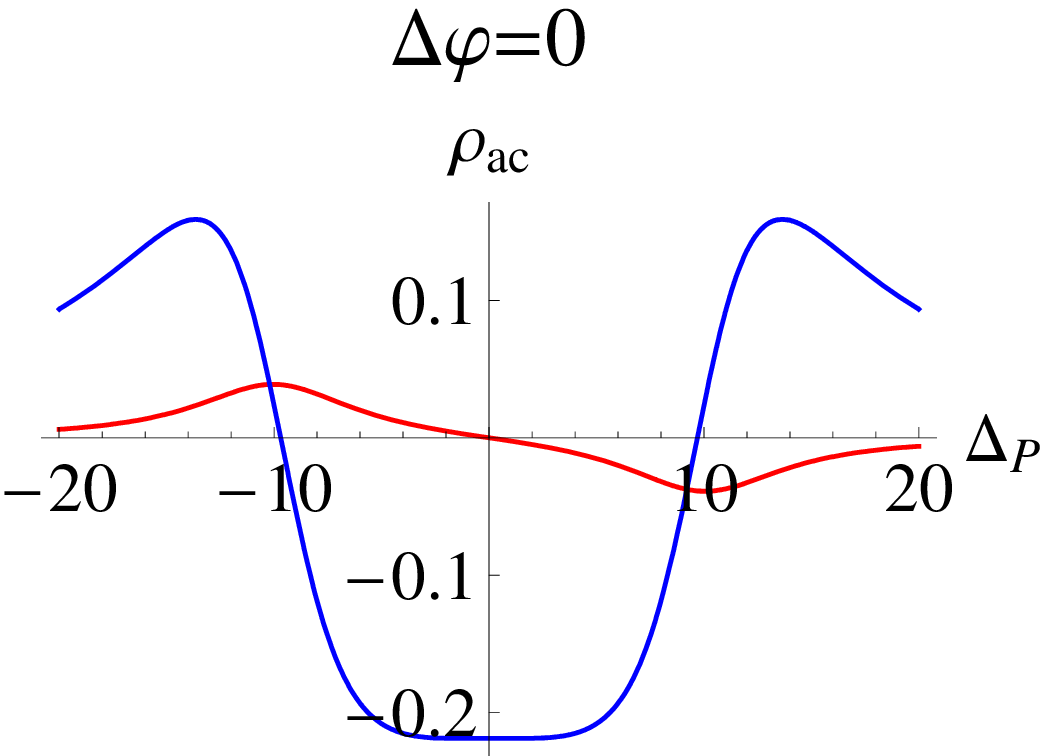}}
\subfigure[]{\label{DFi=05Pi}\includegraphics[scale=0.3]{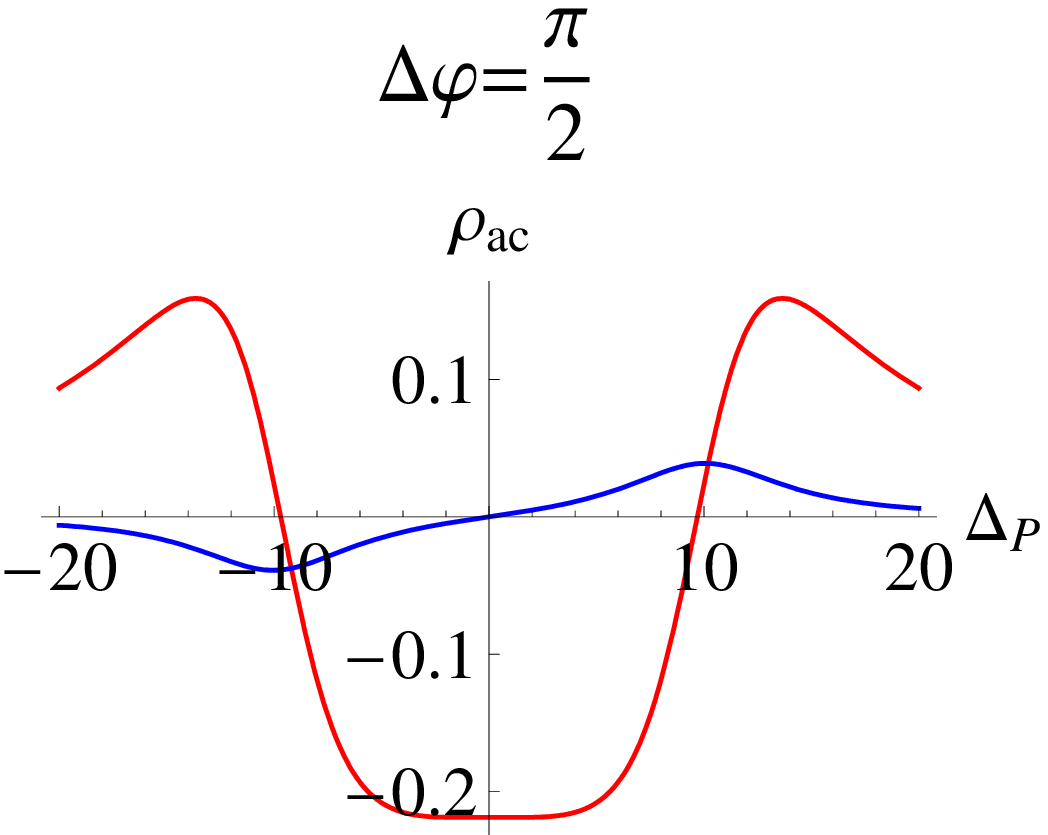}}\\
\subfigure[]{\label{DFi=Pi}\includegraphics[scale=0.3]{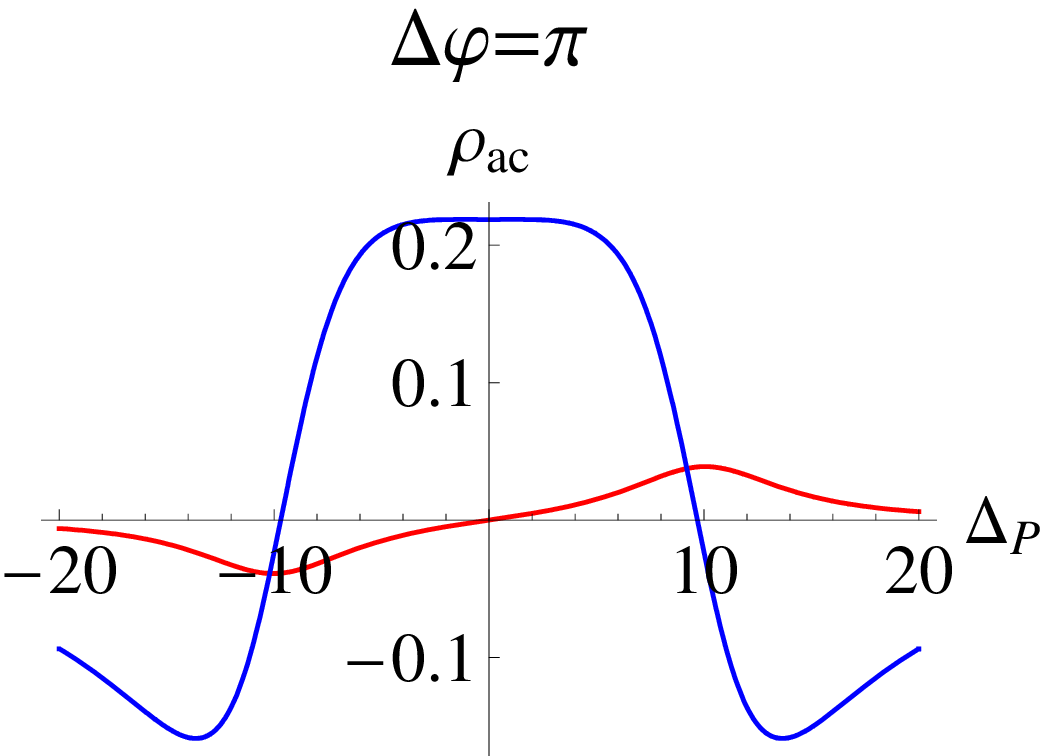}}
%\subfigure[]{\label{r02=-5}\includegraphics[scale=0.23]{Lam-Forb-Spectr-dP=-5.eps}}
\subfigure[]{\label{DFi=15Pi}\includegraphics[scale=0.3]{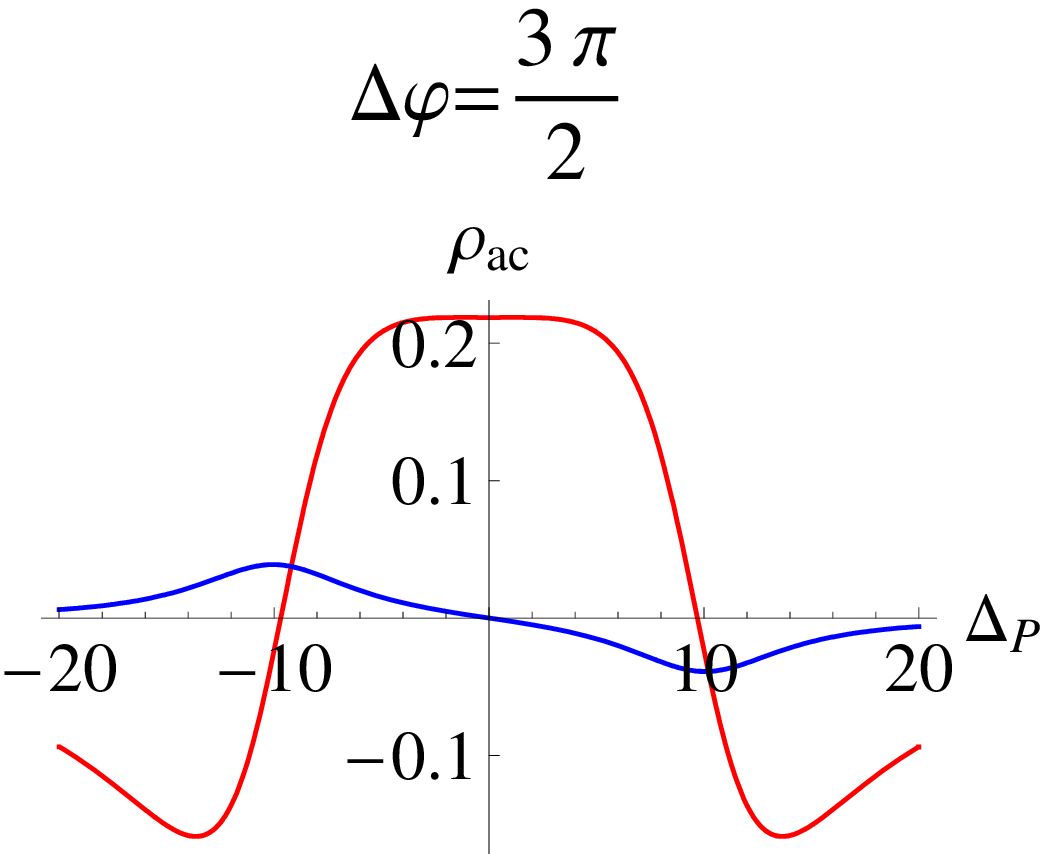}}
%\subfigure[]{\label{r02=5}\includegraphics[scale=0.23]{Lam-Forb-Spectr-dP=5.eps}}\\
%\subfigure[]{\label{r02=10}\includegraphics[scale=0.23]{Lam-Forb-Spectr-dP=10.eps}}
%\subfigure[]{\label{r02=15}\includegraphics[scale=0.23]{Lam-Forb-Spectr-dP=15.eps}}
%\subfigure[]{\label{r02=20}\includegraphics[scale=0.23]{Lam-Forb-Spectr-dP=20.eps}}
\caption{Flattop behavior of gain/absorption (red line) and dispersion (blue line) on the ACSA transition at strong pump
field at various values of relative field phase $\Delta\varphi$. Gain (absorption) is obtained at positive (negative) values. (a) Suppressed refraction index
at $\Delta\varphi=0$, (b) Strong absorption with normal dispersion
at $\Delta\varphi=\pi/2$, (c) Enhanced refraction index at
$\Delta\varphi=\pi$,  (d) Gain without inversion with negative
dispersion slope at $\Delta\varphi=-\pi/2$. Note that the antisymmetric behavior of the gain is accompanied by symmetric behavior of the dispersion at specific phase values, and vise versa. Parameters: $\gamma_{ba}=\gamma_{cb}$, $\Omega_{L}=10\gamma_{cb}$, $\Omega_{P}=3.4\gamma_{cb}$.} \label{Flattop}
\end{figure}

Another intriguing manifestation of quantum interference is obtained
when the probe field is not weak with respect to the driving one. Figure \ref{Flattop-3D} shows flattop behavior of both the absorption/gain and the dispersion at ACSA transition at strong enough probe field. Again the absorption/dispersion regime can be effectively controlled by the relative field phase $\Delta\varphi$. Figure \ref{Flattop} demonstrates four essentially different
cases controlled by the \textit{phase shift} between the probe and
driving fields: (i) reduced refraction index at $\Delta\varphi=0$
(fields in-phase, Fig.\ref{DFi=0}), (ii) strong absorption with normal dispersion at
$\Delta\varphi=\pi/2$ (Fig.\ref{DFi=05Pi}), (iii) strongly enhanced refraction index at
$\Delta\varphi=\pi$ (out-of-phase fields, (Fig.\ref{DFi=Pi})), and (iv) strong gain
without inversion and anomalous dispersion at
$\Delta\varphi=-\pi/2$ (Fig.\ref{DFi=15Pi}). The results shown in Figs.
\ref{DFi=Pi} and \ref{DFi=15Pi} are especially
important as they hint for two interesting potential applications:
laser or/and optical amplifier with a wide spectral range of
operation (see Fig. \ref{DFi=Pi}), and a controllable
atomic dispersion in wide spectral range (see Fig.
\ref{DFi=15Pi}).

\begin{figure*}[htbp]
\centering
\subfigure[]{\label{r02=-20}\includegraphics[scale=0.45]{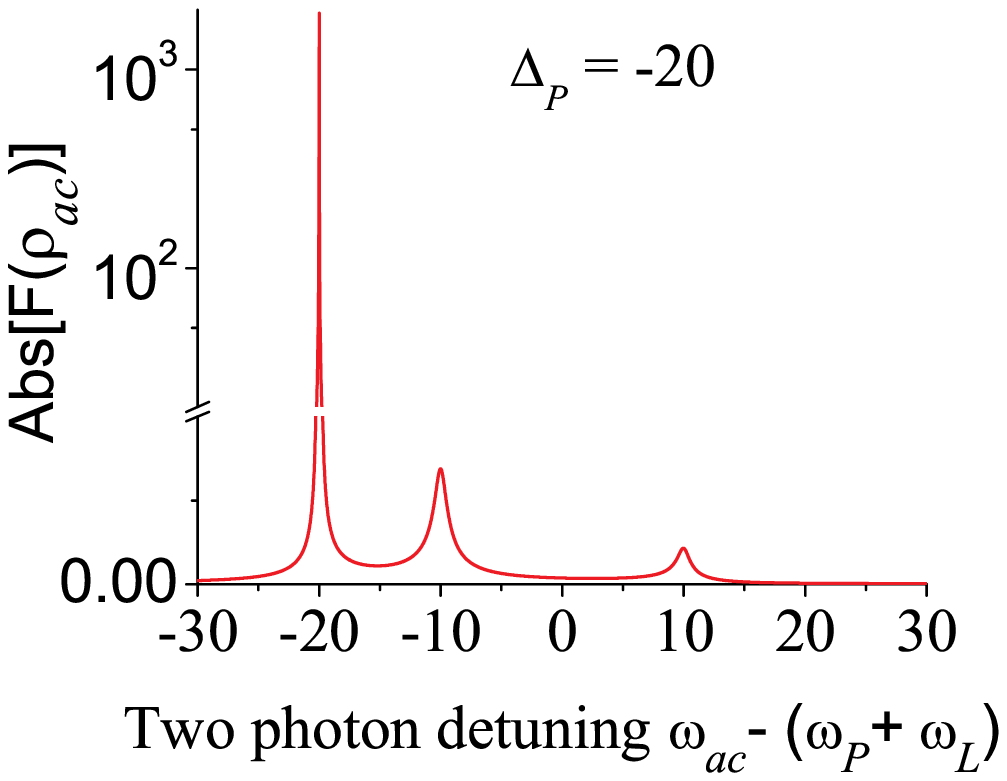}}
\subfigure[]{\label{r02=-15}\includegraphics[scale=0.45]{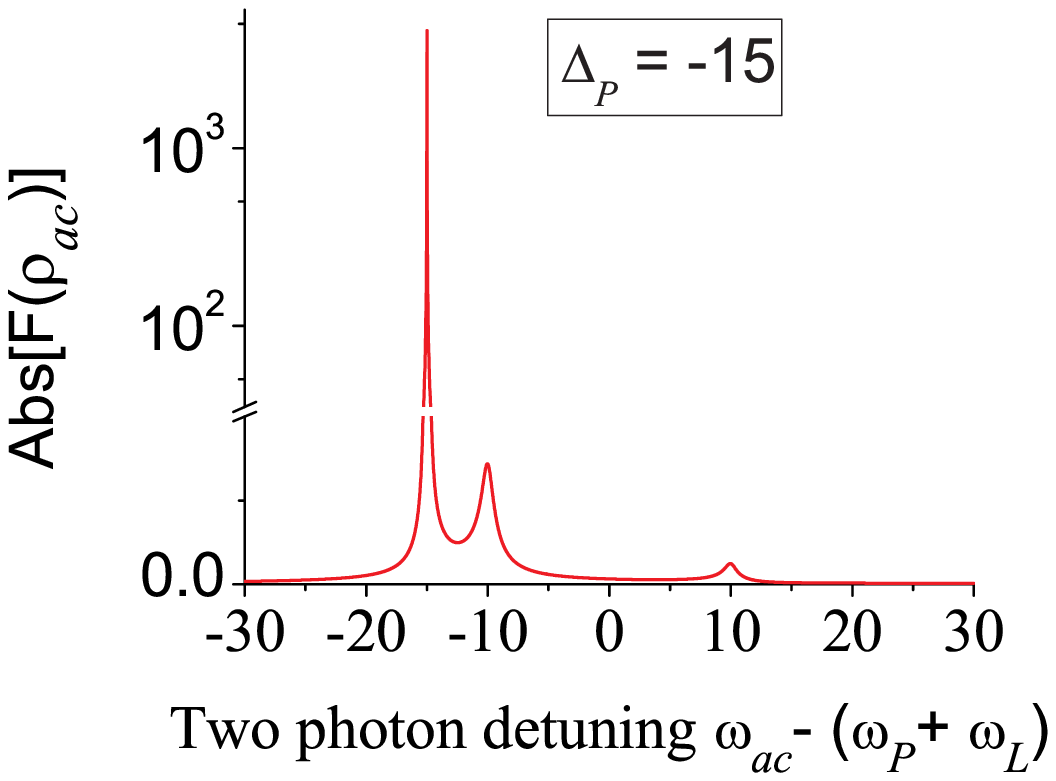}}
\subfigure[]{\label{r02=-10}\includegraphics[scale=0.45]{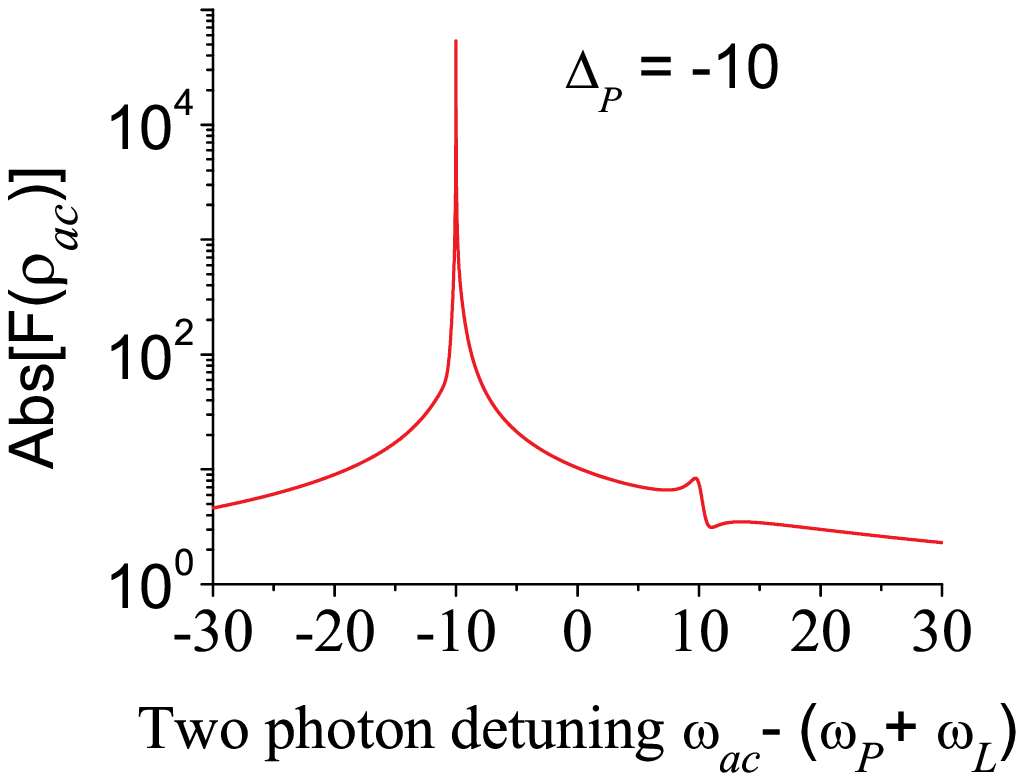}}\\
\subfigure[]{\label{r02=-5}\includegraphics[scale=0.45]{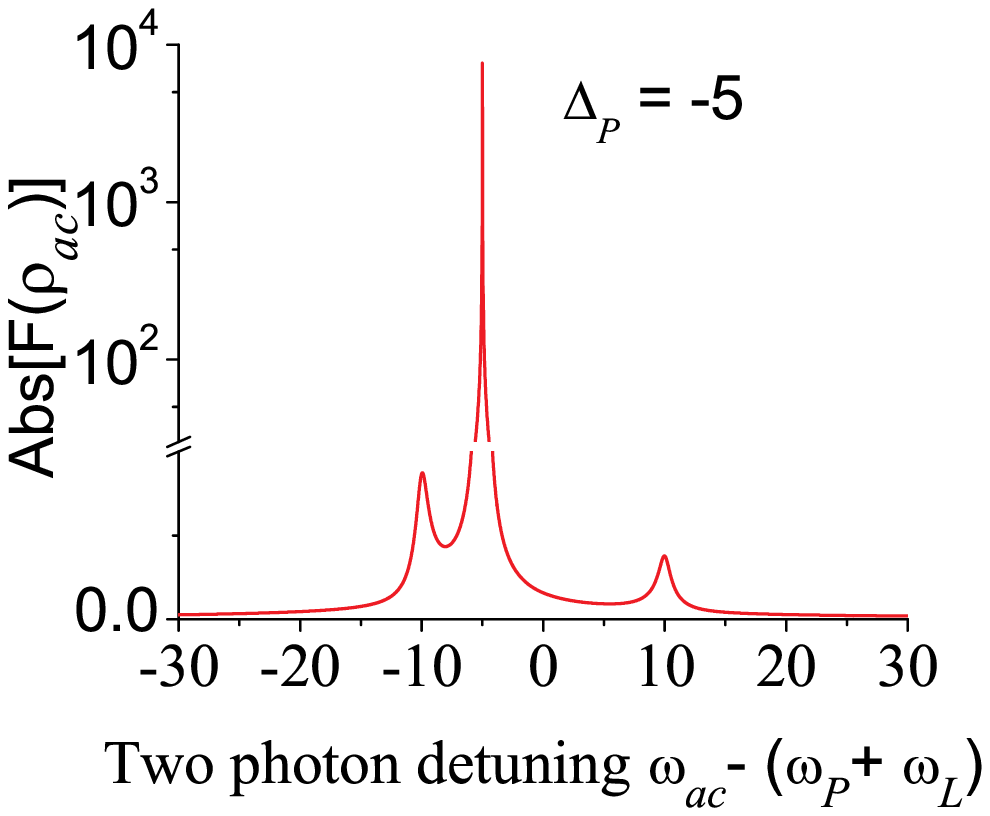}}
\subfigure[]{\label{r02=0}\includegraphics[scale=0.45]{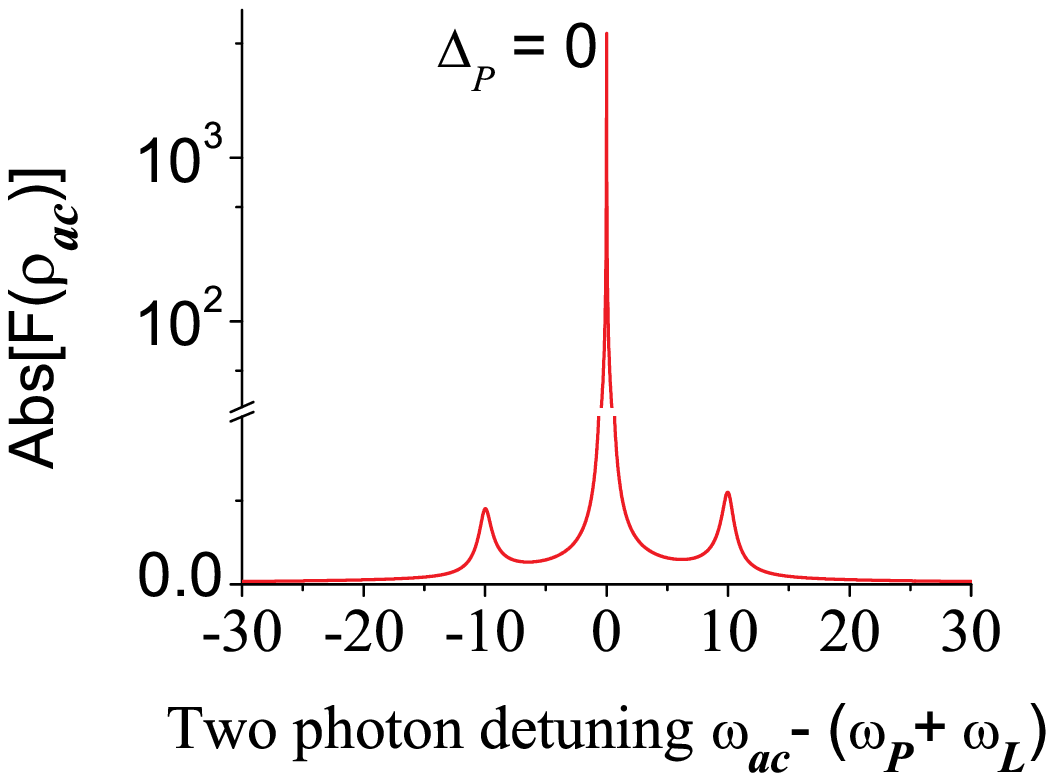}}
\subfigure[]{\label{r02=5}\includegraphics[scale=0.45]{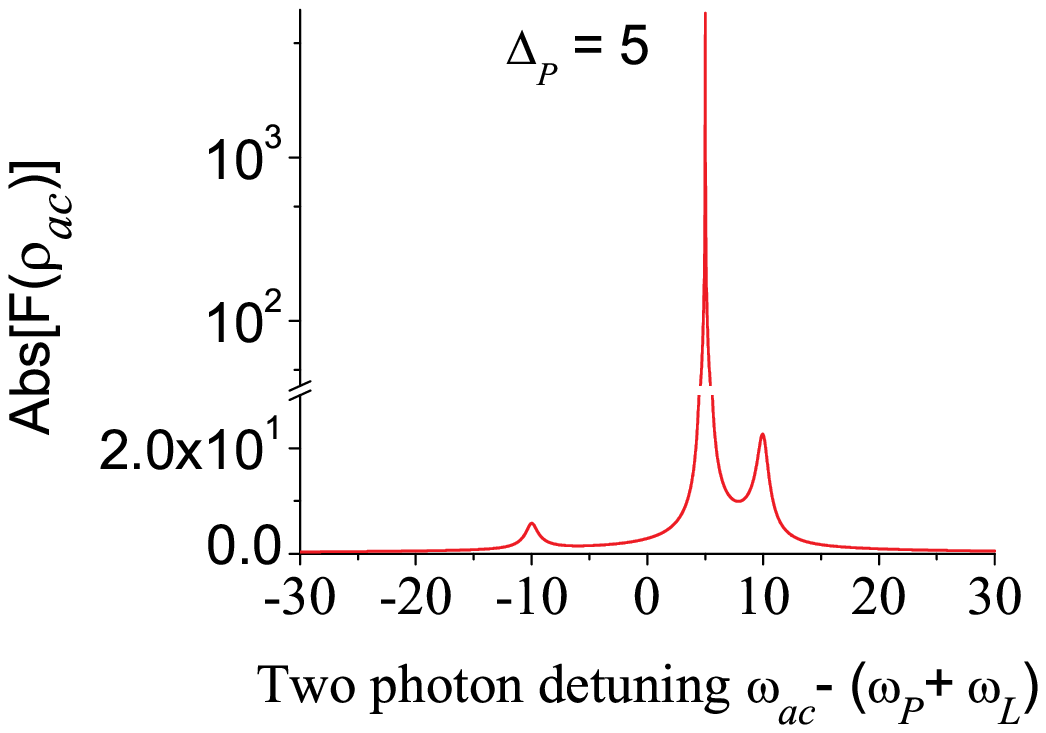}}\\
\subfigure[]{\label{r02=10}\includegraphics[scale=0.45]{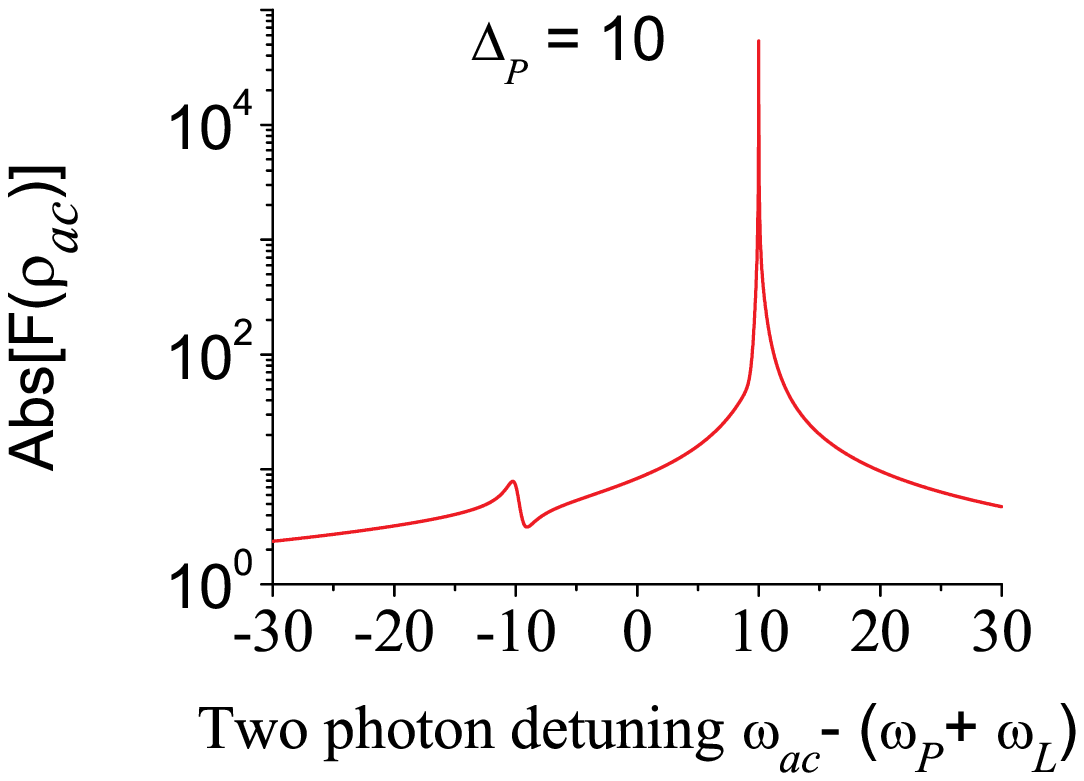}}
\subfigure[]{\label{r02=15}\includegraphics[scale=0.45]{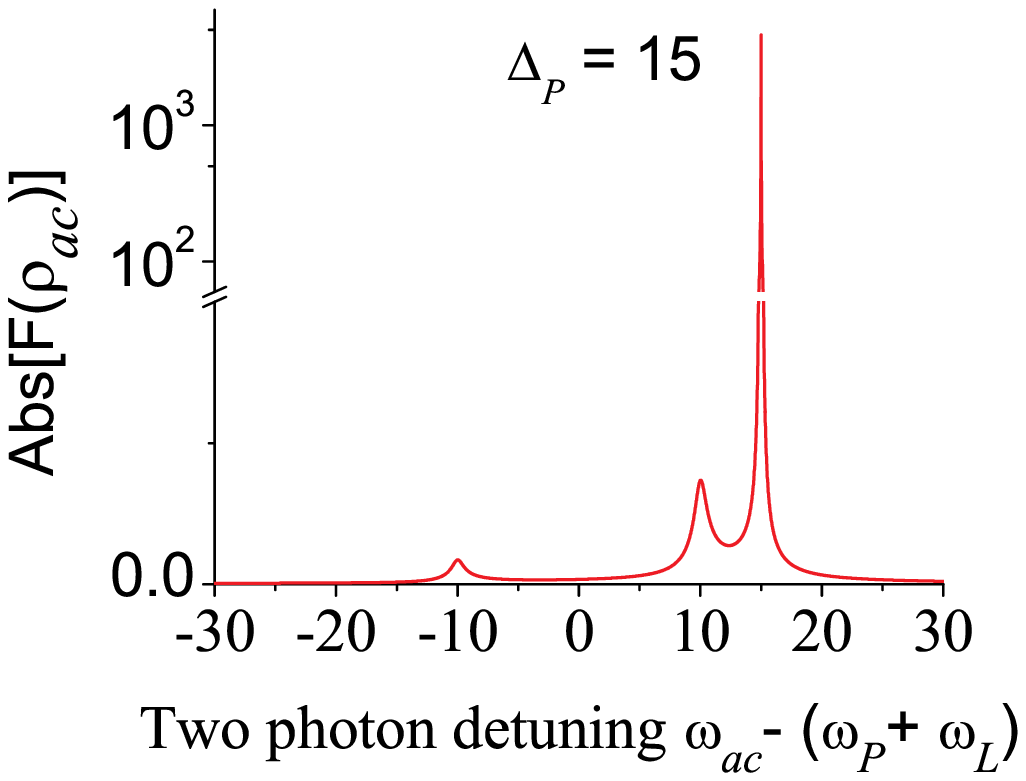}}
\subfigure[]{\label{r02=20}\includegraphics[scale=0.45]{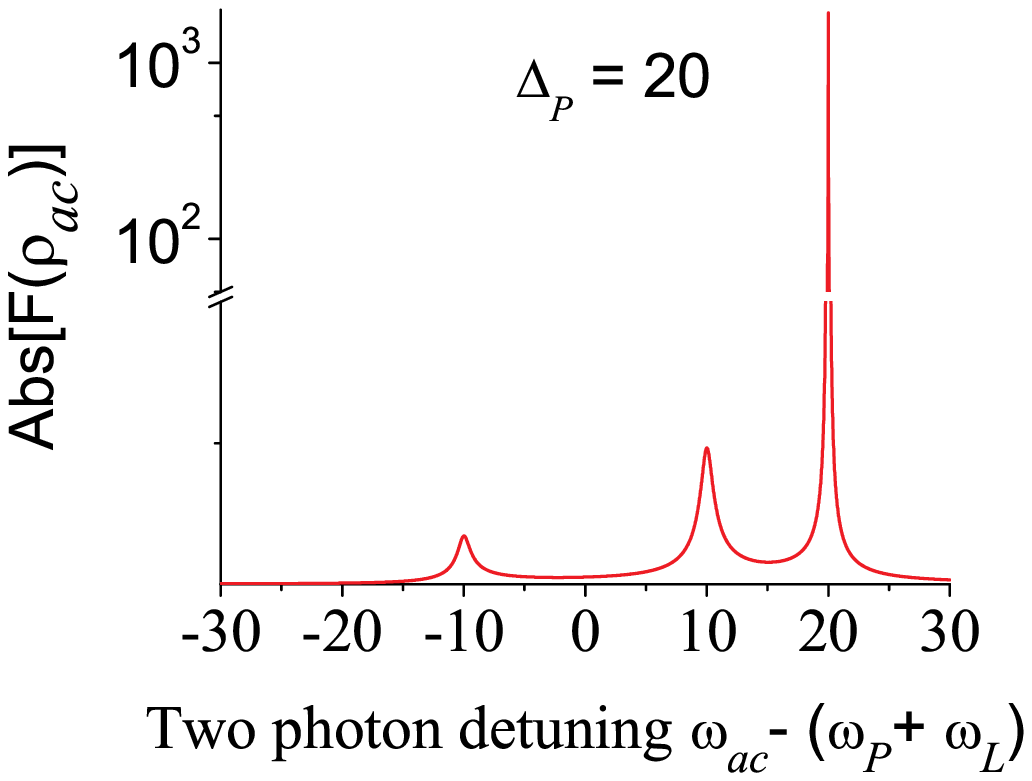}}\caption{Absorption spectrum of
$\left|a\right\rangle\rightarrow\left|c\right\rangle$ ACSA transition. Note sharp strong peak at dressed
two-photon resonance $\Delta_{P}+\Delta_{L}=\pm\Omega_{L}$ (see parts c and g which are plotted in
logarithmic scale).
Parameters: $\Omega_{L}=10, \Omega_{P}=0.37,
\Delta\varphi=0$, $\Delta_{L}=0$, $\Delta_{P}=-20$ (a), $-15$ (b), $-10$ (c), $-5$ (d), $0$ (e), $5$ (f), $10$ (g), $15$ (h), $20$ (i).} \label{ForbidSpetr(deltaP)}
\end{figure*}

\section{Absorption spectrum}
To calculate the absorption spectrum, the master equation (\ref{master}) with time dependent Hamiltonian (\ref{Hamiltonian}) has been solved numerically and then Fourier transform has been taken. Figure \ref{ForbidSpetr(deltaP)} plots the spectrum
of the coherence $\rho_{ac}$ for various values of the probe
detuning $\Delta_{P}$. As one can see, in addition to two resonances
at Rabi-shifted transition frequencies $\omega_{ac}\pm\Omega_{L}$, a
very sharp peak appears at $\omega_{ac}+\Delta_{P}+\Delta_{L}$. This
peak becomes especially strong by a few orders of magnitude, at the
''dressed'' two-photon resonance, i.e. at
$\Delta_{P}+\Delta_{L}=\pm\Omega_{L}$ (see Figs.
\ref{r02=-10} and \ref{r02=10}).
Interestingly, the peaks have extremely narrow line widths. We have found numerically that
increasing accuracy of digital Fourier transform gives rise to
vanishing line width, i.e. ideally, if one could perform Fourier
transform with infinite accuracy, the peaks would look like
$\delta$-function. This can be explained by the absence of
mechanisms of line-broadening on the ACSA transition in this
model. Indeed, there are neither dissipative processes which could
contribute to line width, nor the field broadening as no field is
applied to this transition. These will, of course, give a finite
line width.

Sharp strong resonance in the spectrum of
$\left|a\right\rangle\rightarrow\left|c\right\rangle$ transition
means that an efficient lasing is possible on this transition with
extremely narrow spectral line. Moreover, such a laser would be
tunable since its frequency can be controlled by varying both the
frequency of either the driving or the probe laser, and by the
intensity of the driving laser. Features of dark states and EIT are also obtained.

\section{Conclusions}
The ladder configuration of a three level system interacting with two laser fields has been studied. We have shown that strong coherence is established on the  $\left|a\right\rangle\rightarrow\left|c\right\rangle$ transition that originally was electric-dipole forbidden due to selection rules. However, the presence of two laser fields brakes the spherical symmetry and makes this transition dynamically allowed. Calculation of the coherences shows that ACSA (AC Stark Allowed) transition can exhibit such quantum interference related phenomena as gain without inversion, EIT, enhanced refraction etc. It is demonstrated that absorption/dispersion properties of ACSA transition are phase sensitive as they depend upon the relative phase of the probe and the drive laser fields and are changed at will. Subnatural linewidth resonances are found on ACSA transition enabling high resolution spectroscopy. This may also involve practical applications in Autler-Townes spectroscopy in atomic \cite{Autler-Townes-Rb} and molecular \cite{Autler-Townes-Na} systems. 

Strong gain without inversion found in a wide spectral range at strong probe laser field, is of particular interest as it opens a new perspective for
creating ultra-short wavelength and X-ray lasers without inversion. An important finding interesting for potential applications is simultaneous strong gain and negative dispersion slope, which means that an incident light may both be amplified and slowed down. This leads to significant changes in the group velocity in this range, in the speed of light etc. An occurrence of flat top dispersion accentuates these effects in wide range of detuning of the probe and the drive fields. These are important in broadband communications, Sagnac effects and other applications.

It should be noted that our results are applicable to the systems
where all three transitions are electric dipole allowed, such as
Ruby and other solid materials, and semiconductor
quantum structures.

We propose to use femtosecond comb for the pump and probe lasers where phase relation among components can be determined. In such experiment one looks for radiation at the forbidden transition frequency.

%\bibliographystyle{unsrt}
%\bibliography{2014}

\end{document}